\documentclass[aip, amsmath,amssymb, reprint,]{revtex4-1}%
\usepackage{graphicx}
\usepackage{bm}
\usepackage{epstopdf}
\usepackage{sidecap}
\usepackage{amsmath}
\usepackage{amsfonts}
\usepackage{amssymb}%
\providecommand{\U}[1]{\protect\rule{.1in}{.1in}}
\begin{document}
\title{Subwavelength electromagnetic diode: one-way response of cascading nonlinear meta-atoms}

\author{Yuancheng Fan}
\affiliation{Department of Physics, Tongji University, Shanghai 200092, People's Republic of China}
\affiliation{Shanghai Key Laboratory of Special Artifical Microstructure Materials and Technology, Shanghai, People's Republic of China}
\author{Jin Han}
\affiliation{Department of Physics, Tongji University, Shanghai 200092, People's Republic of China}
\affiliation{Shanghai Key Laboratory of Special Artifical Microstructure Materials and Technology, Shanghai, People's Republic of China}
\author{Zeyong Wei}
\affiliation{Department of Physics, Tongji University, Shanghai 200092, People's Republic of China}
\affiliation{Shanghai Key Laboratory of Special Artifical Microstructure Materials and Technology, Shanghai, People's Republic of China}
\author{Chao Wu}
\affiliation{Department of Physics, Tongji University, Shanghai 200092, People's Republic of China}
\affiliation{Shanghai Key Laboratory of Special Artifical Microstructure Materials and Technology, Shanghai, People's Republic of China}
\author{Yang Cao}
\affiliation{Department of Physics, Tongji University, Shanghai 200092, People's Republic of China}
\affiliation{Shanghai Key Laboratory of Special Artifical Microstructure Materials and Technology, Shanghai, People's Republic of China}
\author{Xing Yu}
\affiliation{Department of Physics, Tongji University, Shanghai 200092, People's Republic of China}
\affiliation{Shanghai Key Laboratory of Special Artifical Microstructure Materials and Technology, Shanghai, People's Republic of China}
\author{Hongqiang Li}
\email{hqlee@tongji.edu.cn}
\affiliation{Department of Physics, Tongji University, Shanghai 200092, People's Republic of China}
\affiliation{Shanghai Key Laboratory of Special Artifical Microstructure Materials and Technology, Shanghai, People's Republic of China}

\begin{abstract}
We propose a scheme for realizing subwavelength electromagnetic diode by employing cascading nonlinear meta-atoms. One-way response is demonstrated on a microwave transmission line comprising of three metallic ring resonators acting as meta-atoms and a varactor as the nonlinear medium inclusion. Experiments show that our implementation can operate simultaneously as forward diode and backward diode at different frequencies. A transmission contrast of up to 14.7dB was achieved between forward and backward transmission. Subwavelength size of our diode should be useful for miniaturization of integrated optical nanocircuits.

\end{abstract}
\maketitle
Time-reversal invariance means that the laws of physics have the same mathematical form even when time is reversed. In optics, this means that light propagation in the forward direction is the same as in the backward direction. While most optical systems are time-reversal invariant, nonreciprocal optical systems are of fundamental interest in optical physics because of their one-way nature. One-way electromagnetic (EM) transmission, or the optical diode effect, has been extensively investigated in nonlinear photonic crystals and optical waveguides\cite{1,2,3,4,5,6,7,8,9,10,11}, for instance, by tuning the band edge\cite{1,2} or defect mode\cite{5}.

With the rapid development of fabrication technology, a new era of subwavelength nanophotonics is approaching. The realization of one-way transmission functionalities within a subwavelength volume is indispensable for all-optical logic components in integrated optical nanocircuits or metactronics\cite{12}. Metamaterials refer to artificial materials with properties unattainable in nature\cite{13,14,15,16}. These artificial materials usually consist of subwavelength-sized metallic resonant building blocks as meta-atoms\cite{17,18}, which yield electric or/and magnetic responses. The local resonances of the meta-atoms and mutual coupling among them enable the amplification of evanescent waves at a subwavelength scale. As such, they can produce strong EM nonlinearity with assistance of strong localization and confinement of the EM field\cite{19,20,21,22,23,24,25,26}. In this paper, we demonstrate a meta-atomic system that can operate as a subwavelength electromagnetic diode (SED). We employ a microwave transmission line comprising of three metallic ring resonators and nonlinear medium inclusion to conceptually demonstrate the one-way response of cascading nonlinear meta-atoms for SED by properly harvesting the antisymmetric mode of meta-atoms under near-field excitation. Our experiments show that the SED can operate simultaneously as a forward diode and a backward diode at different frequencies. A contrast ratio of up to 14.7dB is achieved between forward and backward transmission. The SED sample is only about one fifth of the operational free-space wavelength.\begin{figure}[pb]
\includegraphics[width=8.0cm
]{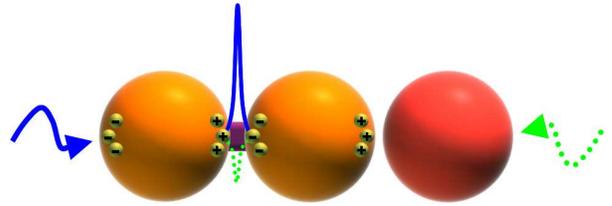}\caption{Schematic of a cascading nonlinear meta-atomic model and local electric fields
enhanced on the nonlinear medium inside the gap between the two meta-atoms on the left under forward
(solid) and backward (dotted) excitation.}%
\end{figure}

Figure 1 schematically illustrates our theoretical model of cascading nonlinear meta-atoms. The gap between the two identical meta-atoms (orange) on the left is filled with nonlinear medium (purple bar). In principle, there exists an antisymmetric mode with respect to two identical meta-atoms in close proximity, for example, a pair of closely spaced metal particles\cite{27}. The antisymmetric mode is established when the surface charges on identical meta-atoms are oscillated out of phase at on-resonance condition, giving rise to a maximized voltage gradient between the meta-atoms. Given that the gap is very narrow, the local electric field (voltage gradient) between the meta-atoms can be several orders larger than the incident field. A third meta-atom (red) on the right will introduce spatial asymmetry to the model system along the direction of the aligned meta-atoms. The induced surface charge density on the meta-atoms will differ depending on the side from which the excitation waves approach, as will the strength of the local electric field inside the gap. The arrows and curves in Fig. 1 denote forward and backward excitation and the corresponding local electric fields within the gap. When the antisymmetric mode is excited, the strength of local electric fields inside the gap can be very large for incidence from the left (blue), causing a giant nonlinear response that is tunable by the input power as well as the gap size, whereas it is small for incidence with same amplitude from the right (green). As a consequence, the asymmetric nonlinear feedback gives rise to a one-way response.

Photograph of our sample model is shown in Fig. 2(a). Three closely spaced metallic ring resonators (MRRs) are evanescently coupled on a transmission line and act as meta-atoms. A silicon hyperabrupt varactor (Infineon BBY52) acting as nonlinear medium is loaded between two identical rectangular MRRs. The sample is fabricated and deposited on a printed circuit board (PCB), which has a metal sheet attached to the back. The PCB substrate is $0.5mm$ thick and has a dielectric constant of $\epsilon_{r}=2.2$. The metal strip is $0.5mm$ wide, the width and length of the two identical rectangular MRRs are $w_{1}=2.5mm$ and $l_{1}=15.5mm$, and those of the third square one are $w_{2}=10.5mm$ and $l_{2}=9.5mm$. The gaps between adjacent MRRs are $g_{1}=0.8mm$ and $g_{2}=0.5mm$ in size. The total length of our cascading meta-atomic model is $18.3mm$. We perform full wave simulations with a finite integration technique (FIT) based electromagnetic solver (CST Microwave Studio) for all the numerical calculations. In our experiments, an Agilent 83017A amplifier and an HP 8493C attenuator are adopted at the input and output ports of an Agilent N5244A vector network analyzer for power control, transmission spectra in forward and backward directions are measured by reversing the samples.\begin{figure}[pb]
\includegraphics[width=8.0cm]{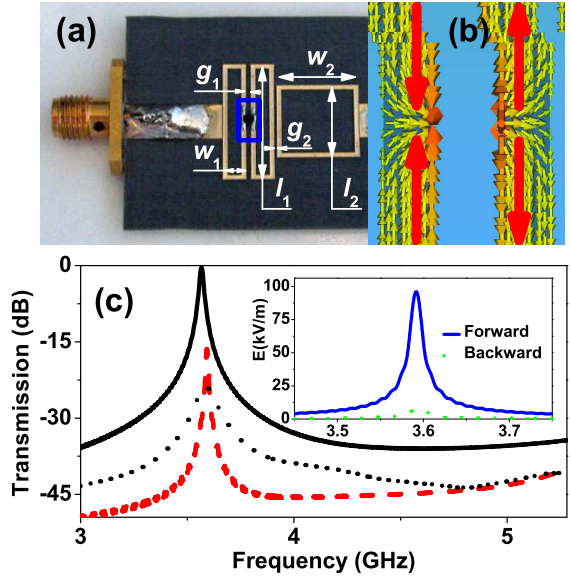}\caption{(a) Photograph of our sample cascading meta-atomic diode.
Geometric parameters are denoted by white letters. (b) Chart of surface currents at 3.56GHz
on the metallic strips inside the solid frame on the photograph. (c) Transmission spectra of the sample
diode (dashed: calculated) and the comparison model (solid: calculated, dotted: measured).
Inset illustrates strength of the local electric field on the capacitor
as a function of frequency under forward (solid) and backward (dotted) excitation.
Input power fed through the transmission line is fixed at $1V/m$.}%
\end{figure}

As a comparative study, we also fabricate a sample of two-resonator system without the third square MRR. We Figure 2(c) presents the calculated and measured transmission spectra of the two-resonator sample (black solid and dotted lines, respectively). The peak position measured at 3.56GHz agrees well with calculations in which a fixed value of $2.5pF$ is assigned to the capacitor. The peak can also be observed in the calculated [red dashed line in Fig. 2(c)] and measured [see Fig. 3(a)] spectra of our sample diode. As Fig. 2(b) shows, surface currents at the peak frequency of 3.56GHz, strongly localized around the gap, oscillate in antiphase on the two rectangular MRRs. This is rightly a picture of the antisymmetric mode [with red arrows in Fig. 2(b) denoting the anti-aligned surface currents oscillated on the adjacent MRRs]. The voltage gradient inside the gap, i.e., the strength of the local electric field on the capacitor, is calculated as a function of frequency for fixed input power of $1V/m$. The inset of Fig. 2(c) shows that the voltage gradient reaches a maximum of about $100kV/m$ for forward excitation (from left to right) as shown by the solid blue curve, whereas it is only about $6kV/m$ for backward excitation (from right to left), as shown by the green dotted line. The large contrast in local fields with respect to different excitation direction demonstrates a substantial nonreciprocal optical response owing to the field enhancement of the antisymmetric mode.\begin{figure}[ptb]
\includegraphics[width=8.6cm
]{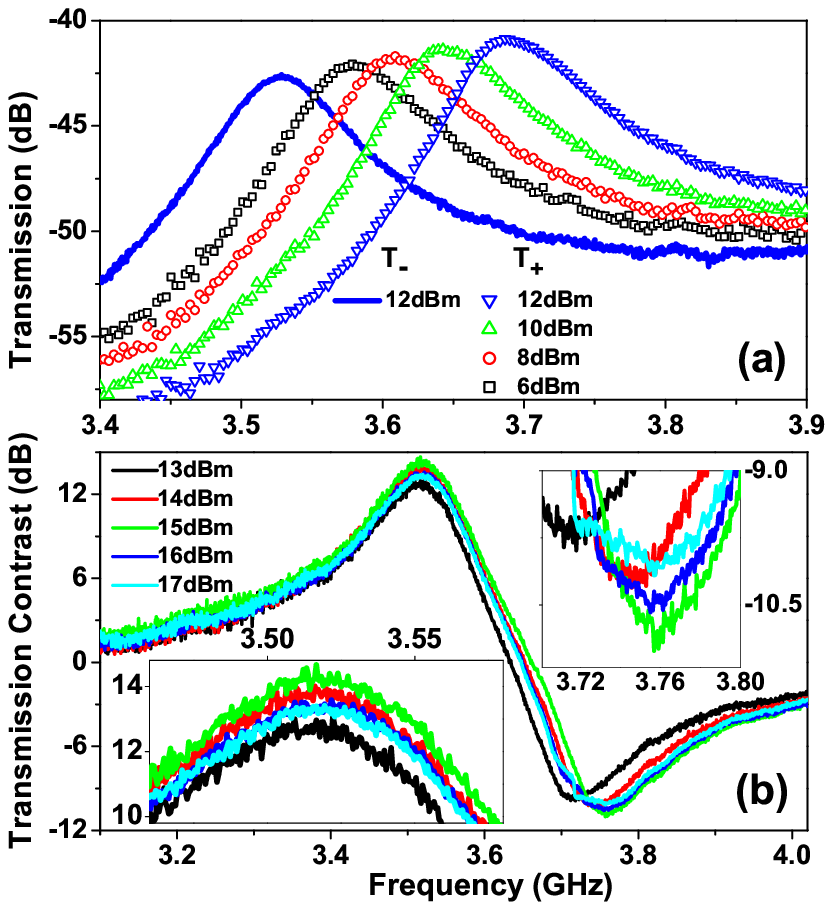}\caption{(a) Backward and forward transmission spectra $T_{-}\left(  f\right)  $
and $T_{+}\left(  f\right)  $ at input powers ranging from $6dBm$ to $12dBm$.
(b) Transmission difference $\Delta T=T_{-}-T_{+}$ at input powers ranging from $13dBm$ to $17dBm$.
Insets show peaks around 3.52GHz and dips around 3.75GHz.}%
\end{figure}

The backward and forward transmission spectra $T_{-}\left(f\right)$ and $T_{+}\left(f\right)$ are measured in the range of $-10\thicksim25dBm$ of input power. With increasing input power, the varactor will undergo a much higher voltage drop under forward excitation than under backward excitation. Figure 3(a) shows the spectra in the frequency range of .4$\thicksim$3.9GHz for input powers ranging from 6dBm to 12dBm. The frequency shift of the antisymmetric mode is very sensitive to the input power of forward excitation [see dots in Fig. 3(a)]. The peak around 3.58GHz at input power of $6dBm$ shifts to 3.69GHz at input power of $12dBm$, whereas the resonant peak is almost fixed at 3.52GHz under backward excitation at different input power levels [for clarity, only the backward transmission spectrum at input power of $12dBm$ is shown, see the solid line in Fig. 3(a)]. Thus, the transmission is nonreciprocal in this frequency range, as predicted for SED. The feature can be visualized more clearly in the spectra of log-scaled transmission difference . The peaks around 3.52GHz and the dips around 3.75GHz in Fig. 3(b) explicitly indicate that our cascading meta-atomic structure can operate simultaneously as a backward diode and a forward diode around these two frequencies.\begin{figure}[ptb]
\includegraphics[width=8.0cm]{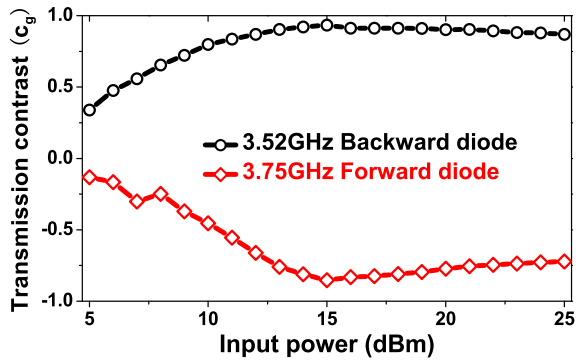}\caption{Contrast ratio
$c_{g}$ of forward and backward transmission at 3.52GHz and 3.75GHz versus input power.}%
\end{figure}

The SED performances at 3.52GHz and 3.75GHz can be evaluated by the contrast ratio of forward and backward transmission magnitudes, defined as $c_{g}=\left(T_{-}-T_{+}\right)/\left(T_{-}+T_{+}\right)$\cite{4}. The ratio at 3.52 GHz, shown as a dotted line with circles in Fig. 4, is a monotonic increasing function about the input power until it reaches a saturation value of $0.934$ at $15dBm$. Note that the peak frequency in the forward transmission spectra (see Fig. 3) increases monotonically with increasing input power below $15dBm$. At input powers higher than $15dBm$, the transmission shows bistable behavior. This can be confirmed by checking the evolution of the peaks and dips around 3.52GHz and 3.75GHz, respectively, shown in the insets in Fig. 3(b) for input powers ranging from 13dBm to 17dBm. We also see that the ratio at 3.75GHz reaches a saturation value of -0.853 at input power of $15dBm$. Note that the model SED of $18.3mm$ is only about one fifth of operational free-space wavelength at 3.52GHz and less than a quarter of the wavelength at 3.75GHz.

In summary, a cascading meta-atomic SED is proposed and implemented on a microwave transmission line. Three MRRs in tandem loaded with one varactor constitute a system of cascading nonlinear meta-atoms. Our experiments prove that the sample system can simultaneously operate as a backward diode and a forward diode in two different frequency ranges. The local electric field ($100kV/m$) between the meta-atoms is enhanced to about five orders of magnitude larger than the incident field ($1V/m$). Strong localization and confinement of EM fields in a subwavelength volume derived from the antisymmetric mode is instrumental for the realization of SED at low working power. Our findings are beneficial for miniaturization of nonlinear optical components for metamaterial-inspired optical nanocircuits \cite{12}.

This work was supported by NSFC (No. 10974144, 60674778), the National 863 Program of China (No. 2006AA03Z407),NCET (07-0621), STCSM and SHEDF (No. 06SG24).

\end{document}